\documentstyle[12pt]{article}
\def\np    { Nucl. Phys. }
\def\pr    { Phys. Rev. }
\def\pl    { Phys. Lett. }

\def\prl   { Phys. Rev. Lett. }

\def\ptp   { Prog. Theor. Phys. }
\def\del {\partial}
\def\delLambda {\frac{\del}{\del\Lambda}}
\def\vec#1 {\mbox{\boldmath $#1$}}
\def\half {{1 \over 2}}
\def\Tr {\mbox{Tr}}
\def\calH {{\cal H}}
\def\calD {{\cal D}}
\def\esprime { s^{\prime} }
\def\esdbprime { s^{\prime\prime} }

\def\&{&\!\!\!\!\!\!\!\! &}

\def\nn{\nonumber}

\def\bigquad{\qquad\qquad\qquad\qquad\qquad\qquad}
\def\nspace { \!\!\!\!\!\!\!\!\!\!\!\!\!\!\!\!\!\!\!\!\!\!\!\! }
\renewenvironment{thebibliography}{\pagebreak[3]\par\vspace{0.6em}
\begin{flushleft}{\large \bf References}\end{flushleft}
\vspace{-1.0em}

\begin{enumerate}\if@twocolumn\baselineskip=0.6em\itemsep -0.2em
\else\itemsep -0.2em\fi\labelsep 0.1em}{\end{enumerate}}

\begin{document}
\baselineskip=0.65cm





\begin{titlepage}

    \begin{normalsize}
     \begin{flushright}
                 NSF-ITP-99-128 \\
                 hep-th/9910256 \\
                 October 1999
     \end{flushright}
    \end{normalsize}
    \begin{LARGE}
       \vspace{1cm}
       \begin{center}
        Exact Renormalization Group \\
          and \\
        Loop Equation \\
       \end{center}
    \end{LARGE}
  \vspace{5mm}

\begin{center}
           Shinji Hirano
           \footnote{E-mail address:
              hirano@itp.ucsb.edu}   \\
      \vspace{4mm}
        {\it Institute for Theoretical Physics}, \\
        {\it University of California},\\
        {\it Santa Barbara, CA 93106, USA}\\
      \vspace{1cm}

    \begin{large} ABSTRACT \end{large}
        \par
\end{center} 
\begin{quote}
 \begin{normalsize}
We propose a gauge invariant formulation of the exact 
renormalization group equation for nonsupersymmetric pure 
$U(N)$ Yang-Mills theory, based on the construction by Tim Morris. In fact 
we show that our renormalization group equation amounts to a regularized 
version of the loop equation, thereby providing a direct relation between 
the exact renormalization group and the Schwinger-Dyson equations.
We also discuss a possible implication of our formulation to the 
holographic correspondence of the bulk gravity and 
the boundary gauge theory.

 \end{normalsize}
\end{quote}

\end{titlepage}
\vfil\eject





\section{Introduction}

The discovery of certain limits in string theory and M-theory
can probably provide a reasoning for the dualities between quantum 
gravity and gauge theory. The celebrated examples are the light like 
limit \cite{SS} of M-theory, which at least literally justifies the 
Matrix conjecture \cite{BFSS}, and the near horizon limits 
\cite{Maldacena,IMSY} of D-branes and M-branes \cite{TASI} that 
led us to recent excitement of the AdS/CFT duality 
\cite{Maldacena, GKP,Witten}. Both the Matrix conjecture and the 
AdS/CFT duality intimately originates from the old $s$-$t$ 
channel duality in open string loop amplitudes. It is, however, quite 
noteworthy that the discovery of the limits of \cite{SS,Maldacena,IMSY} 
elaborated and complemented our intuitive but naive anticipation of 
a dual description of quantum gravity by a certain gauge theory, which 
may be based on the above-mentioned duality of open and closed strings.

The IR/UV relation, pointed out in \cite{SW,PP}, itself is not 
surprising from the viewpoint of the $s$-$t$ channel duality. But 
a rather conceptual payoff of this relation in the near horizon limit 
seems quite remarkable. In particular the identification of the radial 
coordinate $U$ in the near horizon geometry with the energy or the 
cutoff scale $\Lambda$ of the gauge theory clarifies the holographic 
nature of the AdS/CFT duality, in which it is quite plausibly assumed 
that the gauge theory contains only one degree of freedom per cell of 
the cutoff size.\footnote{There is in fact a subtlety in the relation 
between the radial coordinate $U$ and the cutoff scale $\Lambda$ \cite{PP}. 
The existence of two distinct relations is emphasized there. In both 
cases, however, there is a universal property that the cutoff scale 
$\Lambda$ increases with the radial scale $U$ up to dimensions 
$5$ of the boundary gauge theory.}
 This postulate and the resultant holographic property 
actually fit in the basic idea of the Wilsonian renormalization group 
(Wilsonian RG) \cite{WK}.\footnote{The scale invariant theories, of 
course, do not have a nontrivial RG flow. We would, however, like to 
emphasize this point, suppose that the argument in \cite{SW} can be 
applied to more general nonconformal cases.} The assumption concerning 
degrees of freedom of the gauge theory is almost assured, if 
one refers to, for example, a lattice regularization. Once we regard 
the cutoff $\Lambda$ of the boundary gauge theory as the radial 
dimension in the bulk space, degrees of freedom of 
the bulk gravity may well be constituted solely from those of the 
boundary gauge theory, and the coarsening procedure or the RG 
transformation gives the dynamics in the interior space, which will 
likely be in general more complicated than that on the boundary. Thus 
we hope in this respect that the RG flow may correspond to the 
holographic mapping of the boundary data.

The motivation of this paper is to invent a formulation, on the gauge 
theory side, that might be a useful setting for discussing this hope.
Since in the AdS/CFT duality gravitational modes in the bulk 
correspond to the gauge invariant operators on the boundary gauge 
theory, a gauge invariant formulation, in which the gauge redundancy 
is eliminated, may be much suited for discussing this sort of duality. 
In this regard we would like to respect a technique of collective field 
theory developed in \cite{JS}, and apply it to a gauge invariant 
formulation of the exact RG equation. 

There exist several papers \cite{AdSRG}, in which the authors 
discussed the RG interpretation of the AdS/CFT duality and 
some of them suggested the equivalence of the equation of motion in 
the AdS space (and its slight generalization) with the RG 
equation.\footnote{S.-J. Rey also bears an idea to develop this line 
of arguments \cite{SJR}.} We hope that our formulation adds a new 
perspective along the line of their arguments.




\section{A Gauge Invariant Formulation of The Exact RG Equation}

Recently Tim Morris \cite{Morris} proposed an exact RG equation for 
the Yang-Mills theory in a manifestly gauge invariant way, by 
constructing it in terms of a gauge invariant variable, a Wilson loop, 
just like the philosophy of the collective field method \cite{JS}.
His formulation is based on an observation inspired from a simple 
derivation \cite{Polchinski} of the exact RG, that the exact RG 
equation is related to a particular field redefinition of the theory. 
For example, in scalar field theories in $D$ dimensions, the exact RG 
equation may be written in a form,
\begin{equation}
\delLambda e^{-S} = \int d^D\vec{x}
    \frac{\delta}{\delta\phi( \vec{x} )}
    \left(\Psi\left[\phi( \vec{x} )\right]e^{-S}\right),
 \label{eqn:ERG}
\end{equation}
where $\Psi\left[\phi( \vec{x} )\right]$ is induced from an 
infinitesimal change of the scalar field, $\phi( \vec{x} )\to
\phi( \vec{x} ) + \delta\Lambda\Psi\left[\phi( \vec{x} )\right]$.
The exact RG equation \cite{Polchinski}, which was employed to prove 
the renormalizability of the $\lambda\phi^4$ theory in four dimensions, 
is obtained by choosing the field redefinition as
\begin{equation}
\Psi\left[\phi( \vec{x} )\right]=\half\int d^D \vec{y}
 \left[\dot{G}_{\Lambda}(\vec{x} -\vec{y} )\frac{\delta S}
 {\delta\phi( \vec{y} )} - 2\left(\dot{G}_{\Lambda}
 \cdot G^{-1}_{\Lambda}\right)(\vec{x} -\vec{y} )
 \phi(\vec{y} )\right],
 \label{eqn:FRD} 
\end{equation}
where $G_{\Lambda}(\vec{x} -\vec{y} )$ is the cutoff propagator of a 
massless scalar, defined by
\begin{equation}
G_{\Lambda}(\vec{x} -\vec{y} )=\int d^D \vec{p} {1 \over p^2}
  e^{i\vec{p} \cdot (\vec{x} -\vec{y} )}K(p^2/\Lambda^2),
 \label{eqn:COP}
\end{equation}
in which $K(p^2/\Lambda^2)$ is a cutoff function that will take the 
value $1$ for $p^2 < \Lambda^2$ and vanish rapidly at infinity. Also 
$\dot{G}_{\Lambda}$ is the derivative of the propagator with respect to 
the cutoff $\Lambda$, i.e., $\dot{G}_{\Lambda}=\delLambda G_{\Lambda}$.

In pure Yang-Mills theory in $D$ dimensions we may write the exact RG 
equation as
\begin{equation}
\delLambda e^{-S} = \Tr\int d^D\vec{x}
    \frac{\delta}{\delta A^{\mu}( \vec{x} )}
    \left(\Psi^{\mu}\left[A^{\mu}( \vec{x} )\right]e^{-S}\right).
 \label{eqn:ERGYM}
\end{equation}
Here we introduced a standard convention, $A^{\mu}( \vec{x} )=T^a A^{\mu}_a 
( \vec{x} )$ and $\frac{\delta}{\delta A^{\mu}( \vec{x} )}=T^a
\frac{\delta}{\delta A^{\mu}_a( \vec{x} )}$.
A straightforward adaptation of the above regularization scheme 
(\ref{eqn:FRD}), however, spoils the gauge symmetry. One way to avoid it is 
to look for some other form of the field redefinition (\ref{eqn:FRD}) 
which maintains the gauge invariance. Indeed one such a choice was found 
in \cite{Morris}, in which a trick, the introduction of a pair of 
Wilson lines into the field redefinition, seemed to play an essential 
role. We make use of this trick, but we propose a somewhat similar but 
different formulation of the gauge invariant exact RG equation. In 
fact our formulation can be directly connected with the loop equation 
\cite{MM}, as is rather different from the one proposed in \cite{Morris}.

Our choice of the field redefinition is 
\begin{eqnarray}
\Psi_{\mu}[A^{\mu}(\vec{x} )]&=&-{1 \over N\Lambda^3}\int d^D \vec{y} 
 \int d^D \vec{p} e^{i\vec{p} \cdot(\vec{x-y} )}
 \left[K^{'}(p^2/\Lambda^2)\Phi[\Gamma_{\vec{x} \vec{y} }]
 \frac{\delta S}{\delta A^{\mu}(\vec{y} )}
 \Phi^{-1}[\Gamma_{\vec{x} \vec{y} }]\right. \nn\\
 && +2\left. K^{'}(p^2/\Lambda^2)K^{-1}(p^2/\Lambda^2)
 \Phi[\Gamma_{\vec{x} \vec{y} }]{1 \over 2g_b^2}D^{\nu}F_{\nu\mu}(\vec{y} )
 \Phi^{-1}[\Gamma_{\vec{x} \vec{y} }]\right. \nn\\
 && \left. -K^{'}(p^2/\Lambda^2)\frac{\delta}{\delta A_a^{\mu}
 (\vec{y} )}\left(\Phi[\Gamma_{\vec{x} \vec{y} }]T^a 
 \Phi^{-1}[\Gamma_{\vec{x} \vec{y} }]\right)\right].
 \label{eqn:YMFRD}
\end{eqnarray}
This corresponds to the Fourier expansion of eq.(\ref{eqn:FRD}).
Here $\Phi[\Gamma_{\vec{x} \vec{y} }]$ is an advertised Wilson line, 
and defined by $\Phi[\Gamma_{\vec{x} \vec{y} }]=
Pe^{i\oint_{\Gamma_{\vec{x} \vec{y} }}d\vec{x^{\prime}} \cdot\vec{A} 
(\vec{x^{\prime}} )}$, 
where the contour $\Gamma_{\vec{x} \vec{y} }$ is a line from $\vec{x} $ to 
$\vec{y} $.
Also $K^{'}(x)$ denotes the derivative of $K(x)$, i.e., 
$K^{'}(x)={d \over dx}K(x)$, and $g_b$ is the bare Yang-Mills coupling.
In this form it is not clear whether the exact RG equation 
(\ref{eqn:ERGYM}) is gauge invariant. We will, however, see below 
that it is indeed the case.

Now we can formally integrate the RG equation (\ref{eqn:ERGYM}), and 
it takes the form
\begin{equation}
e^{-S}=e_D^{\calH [A,{\delta \over \delta A};\Lambda]}e^{-S_b},
\label{eqn:intERGYM}
\end{equation}
with the bare action $S_b$ and 
\begin{eqnarray}
\calH [A,{\delta \over \delta A};\Lambda]
 &=& 
 {1 \over 2N}
 \int d^D \vec{x} \int d^D \!\!\vec{y} 
 \int \!\!d^D \vec{p} {1 \over p^2}e^{i\vec{p} \cdot(\vec{x-y} )}
 \frac{\delta}{\delta A^{\mu}_b (\vec{x} )}
 \Biggl[ \nn\\
 &&\left. - K^{'}(p^2/\Lambda^2)
 \Tr\left(T^b \Phi[\Gamma_{\vec{x} \vec{y} }]T^a
 \Phi^{-1}[\Gamma_{\vec{x} \vec{y} }]\right)
 \frac{\delta}{\delta A^{\mu}_a(\vec{y} )}\right. \nn\\
 && +\left. K^{'}(p^2/\Lambda^2)K^{-1}(p^2/\Lambda^2)\Tr\left(T^b
 \Phi[\Gamma_{\vec{x} \vec{y} }]{1 \over g_b^2}D^{\nu}F_{\nu\mu}(\vec{y} )
 \Phi^{-1}[\Gamma_{\vec{x} \vec{y} }]\right)\right. \nn\\
 && \left. -K^{'}(p^2/\Lambda^2)
 \frac{\delta}{\delta A_a^{\mu}
 (\vec{y} )}\Tr\left(T^b \Phi[\Gamma_{\vec{x} \vec{y} }]T^a 
 \Phi^{-1}[\Gamma_{\vec{x} \vec{y} }]\right)\right],
\label{eqn:H}
\end{eqnarray}
where the first functional derivative in the r.h.s. operates passing 
through the square brackets as well. Also $e_D^{\calH [A,{\delta \over \delta A};\Lambda]}$ denotes a Dyson series,\footnote{I would like to thank Marty 
Halpern for pointing out an error in the previous version.}
\begin{eqnarray}
\sum_{n=0}^{\infty}\int_0^{{1 \over \Lambda^2}}d(1/\Lambda_n^2)
  \int_0^{{1 \over \Lambda_n^2}}d(1/\Lambda_{n-1}^2)
  \cdots
  \int_0^{{1 \over \Lambda_1^2}}d(1/\Lambda_0^2)
  \calH [A,{\delta \over \delta A};\Lambda_n]
  \calH [A,{\delta \over \delta A};\Lambda_{n-1}]
  \cdots
  \calH [A,{\delta \over \delta A};\Lambda_0].
 \label{eqn:dyson}
\end{eqnarray}
Next let us consider the generating functional for Wilson loop 
correlators, as we are interested only in correlation functions of 
the gauge invariant operators. The generating 
functional is given by\footnote{In the standard formulation of the 
exact RG, the source $J$ for a local operator is chosen in such a 
way that $J(p)$ is vanishing for higher momentum $p^2 > \Lambda^2$ 
\cite{Polchinski}. But the source $J(C)$ introduced here is the one 
for a non-local operator, and so it is unclear what choice is 
appropriate for the exact RG. We will not make any restrictions 
on the source at this stage, instead it will be constrained by the 
consistency of the exact RG equation, as we will see in the next 
section.} 
\begin{equation}
Z[J]=\int\calD A^{\mu}\exp{\left(-S+\sum_C J(C)W(C)\right)},
 \label{eqn:GNLFNC}
\end{equation}
with the definition of the Wilson loop 
$W(C)={1 \over N}\Tr Pe^{i\oint_C d\vec{x} \cdot \vec{A} (\vec{x} )}$.
Using the integrated expression (\ref{eqn:intERGYM}) of the exact RG 
equation and performing the integration by parts, we can rewrite it as 
\begin{equation}
Z[J]
 =\int\calD A^{\mu}e^{-S_b}\left(e_{\widetilde{D}}^{
 \widetilde{\calH} [W,{\delta \over \delta W};\Lambda]}
 e^{\sum_C J(C)W(C)}\right),
 \label{eqn:RWGNLFNC}
\end{equation}
where $e_{\widetilde{D}}^{
\widetilde{\calH} [W,{\delta \over \delta W};\Lambda]}$ denotes a 
Dyson series in the reverse order to (\ref{eqn:dyson}), i.e., 
\begin{eqnarray}
\!\!\!\!
\sum_{n=0}^{\infty}\!\int_0^{{1 \over \Lambda^2}}\!\!d(1/\Lambda_n^2)
  \!\!\int_0^{{1 \over \Lambda_n^2}}\!\!d(1/\Lambda_{n-1}^2)
  \!\cdots\!\!
  \int_0^{{1 \over \Lambda_1^2}}\!\!d(1/\Lambda_0^2)
  \widetilde{\calH} [W,{\delta \over \delta W};\Lambda_0]
  \widetilde{\calH} [W,{\delta \over \delta W};\Lambda_1]
  \!\cdots\!
  \widetilde{\calH} [W,{\delta \over \delta W};\Lambda_n],
 \label{eqn:revdyson}
\end{eqnarray}
and 
$\widetilde{\calH} [W,{\delta \over \delta W};\Lambda]$ is written 
only in terms of Wilson loops. This shows the manifestation of gauge 
invariance in our formulation. We will relegate the detailed calculation 
of the operator $\widetilde{\calH} [W,{\delta \over \delta W};\Lambda]$ 
to the appendix. It finally comes up with a rather suggestive form which 
looks like the string field Hamiltonian,
\begin{eqnarray}
\widetilde{\calH} [W,{\delta \over \delta W};\Lambda]
&=&
 \half\int d^D \vec{p} {1 \over p^2}\left[ 
K^{'}(p^2/\Lambda^2)\Biggl\{ 
\right. \nn\\
&& \nspace\nspace\left.
 {1 \over N^2}\sum_{C,C^{\prime}}
\int_0^{2\pi}ds \int_0^{2\pi}d\esprime
\left(\vec{\dot{x}} (s)\cdot\vec{\dot{x}} (\esprime)\right)
e^{i\vec{p} \cdot(\vec{x} (\esprime)-\vec{x} (s) )}
W(C_M)\frac{\delta}{\delta W(C^{\prime})}\frac{\delta}{\delta W(C)}
\right. \nn\\
&& \nspace\nspace
\left. + \sum_{C}
\int_0^{2\pi}ds \int_0^{2\pi}d\esprime
\left(\vec{\dot{x}} (s)\cdot\vec{\dot{x}} (\esprime)\right)
e^{i\vec{p} \cdot(\vec{x} (\esprime)-\vec{x} (s) )}
W(C_B)W(C_{{\bar B}})\frac{\delta}{\delta W(C)}
\right\} 
\label{eqn:SFH} \\
&& \nspace\nspace\left. 
-{1 \over g_b^2 N}\int d^D \vec{y} K^{'}(p^2/\Lambda^2)K^{-1}(p^2/\Lambda^2)
\sum_C\int_0^{2\pi}ds e^{i\vec{p} \cdot (\vec{x} (s)-\vec{y} )}
\left. \frac{\delta^2 W(C)}{\delta x(s_0)^2}
 \right|_{\vec{x} (s_0)=\vec{y} }\frac{\delta}{\delta W(C)}
 \right]. \nn
\end{eqnarray}
As explained in the appendix, a loop $C_M$ denotes the merging of two loops 
$C$ and $C^{\prime}$, and is given by a product of four line contours as 
$C_M=C_{\vec{x} (s)\vec{x} (s)}\Gamma_{\vec{x} (s)\vec{x} (\esprime)}
C_{\vec{x} (\esprime)\vec{x} (\esprime)}\Gamma_{\vec{x} (\esprime)
\vec{x} (s)}$. 
Also a pair of loops $C_B$ and $C_{{\bar B}}$ is broken up from a loop 
$C$, and they are expressed as $C_B=\Gamma_{\vec{x} (\esprime)\vec{x} (s)}
C_{\vec{x} (s)\vec{x} (\esprime)}$ and 
$C_{{\bar B}}=C_{\vec{x} (\esprime)\vec{x} (s)}
\Gamma_{\vec{x} (s)\vec{x} (\esprime)}$. Here we distinguished a pair of 
lines $\Gamma_{\vec{x} \vec{y} }$ introduced in the field 
redefinition (\ref{eqn:YMFRD}) from the other lines 
$C_{\vec{x} \vec{y} }$ which are parts of loops $C$ and 
$C^{\prime}$.

We will discuss later a possible implication of this result in the 
holographic correspondence of the bulk gravity and the boundary gauge 
theory.




\section{Loop Equation from The Exact RG Equation}

The Schwinger-Dyson equation is supposed to contain as much information 
as the exact RG equation, in a sense that, if we were to solve either of 
these two equations nonperturbatively, we could in principle obtain 
all the physical information of the quantum field theory. Thus we will 
be expected to have an intrinsic way to translate the Schwinger-Dyson 
equation into the exact RG equation, and vice versa, implicitly or 
explicitly. Actually we need to check if our proposed formulation really 
works, by, say, rederiving a known result obtained from a reliable 
formulation. 
In this respect it turns out interestingly enough that our exact RG 
equation amounts to a regularized version of the loop equation of 
\cite{MM}, while this result is not so surprising and in fact rather 
reasonable, as we just mentioned. 
To see it, let us note that eq.(\ref{eqn:RWGNLFNC}) can be further 
rewritten into the form 
\begin{equation}
Z[J]=e_D^{\widetilde{\calH} [{\delta \over \delta J},J;\Lambda]}
     Z_b[J],
  \label{eqn:Hologram}
\end{equation}
where $Z_b[J]$ is the generating functional of the bare form, i.e., 
\begin{equation}
Z_b[J]=\int\calD A^{\mu}\exp{\left(-S_b +\sum_C J(C)W(C)\right)},
 \label{eqn:bareGNLFNC}
\end{equation}
and $e_D^{\widetilde{\calH} [{\delta \over \delta J},J;\Lambda]}$ is 
a Dyson series in the same order as (\ref{eqn:dyson}).
Also in the operator 
$\widetilde{\calH} [{\delta \over \delta J},J;\Lambda]$ 
the derivatives ${\delta \over \delta J}$'s are ordered on the right 
side of the sources $J$'s.

Now the exact RG equation implies 
\begin{equation}
{d \over d\Lambda}Z[J]=0.
 \label{eqn:LambdaIND}
\end{equation}
This is equivalent to 
\begin{eqnarray}
0&=&\int d^D \vec{p} {1 \over p^2}
\left(\delLambda K(p^2/\Lambda^2)\right)K^{-1}(p^2/\Lambda)
\sum_C J(C)\int_0^{2\pi} ds \Biggl[ \nn\\
&&
{1 \over N^2}\sum_{C^{\prime}}
J(C^{\prime})\int_0^{2\pi}ds^{\prime}
\left(\vec{\dot{x}} (s)\cdot\vec{\dot{x}} (s^{\prime})\right)
K(p^2/\Lambda^2)e^{i\vec{p} \cdot(\vec{x} (s^{\prime})-\vec{x} (s) )}
\frac{\delta Z[J]}{\delta J(C_M)}
 \nn\\
&&  + \int_0^{2\pi}d\esprime
\left(\vec{\dot{x}} (s)\cdot\vec{\dot{x}} (\esprime)\right)
K(p^2/\Lambda^2)e^{i\vec{p} \cdot(\vec{x} (\esprime)-\vec{x} (s) )}
\frac{\delta^2 Z[J]}{\delta J(C_B)\delta J(C_{{\bar B}})}
\label{eqn:ERGLOOP} \\
&& \left. 
-{1 \over g_b^2 N}\int d^D \vec{y} 
e^{i\vec{p} \cdot (\vec{x} (s)-\vec{y} )}
\left. \frac{\delta^2}{\delta x(s_0)^2}
\frac{\delta Z[J]}{\delta J(C)}
 \right|_{\vec{x} (s_0)=\vec{y} }
 \right]. \nn
\end{eqnarray}
Therefore we may recognize that the quantity in the square bracket is 
vanishing, and then integrating over the momentum $\vec{p} $, we come 
up to the loop equation with a regularization, 
\begin{eqnarray}
{1 \over g_b^2 N}
\frac{\delta^2}{\delta x(s)^2}
\frac{\delta Z[J]}{\delta J(C)} 
\nspace &&\nn\\
&&\nspace =
{1 \over N^2}\sum_{C^{\prime}}
J(C^{\prime})\int_0^{2\pi}ds^{\prime}
\left(\vec{\dot{x}} (s)\cdot\vec{\dot{x}} (s^{\prime})\right)
\int d^D\vec{p} 
K(p^2/\Lambda^2)e^{i\vec{p} \cdot(\vec{x} (s^{\prime})-\vec{x} (s) )}
\frac{\delta Z[J]}{\delta J(C_M)}
 \nn\\
&& \nspace + \int_0^{2\pi}d\esprime
\left(\vec{\dot{x}} (s)\cdot\vec{\dot{x}} (\esprime)\right)
\int d^D\vec{p} 
K(p^2/\Lambda^2)e^{i\vec{p} \cdot(\vec{x} (\esprime)-\vec{x} (s) )}
\frac{\delta^2 Z[J]}{\delta J(C_B)\delta J(C_{{\bar B}})}.
 \label{eqn:LOOPEQ}
\end{eqnarray}
Note that interestingly the cutoff function $K(p^2/\Lambda^2)$ enters 
in an expected way. In fact $\int d^D\vec{p} 
K(p^2/\Lambda^2)e^{i\vec{p} \cdot(\vec{x} (\esprime)-\vec{x} (s) )}$ 
can be thought of as a smeared $\delta$-function, which is necessary 
to regularize the loop equation. Also loops are ordinarily closed by 
this $\delta$-function, and so the smearing of the $\delta$-function 
might undergo a potential breakdown of the gauge symmetry. It is, however, 
obvious from our explicit computation that loops are closed in spite of 
the smearing of the $\delta$-function in our formulation, due to a pair 
of Wilson lines introduced in the field redefinition 
(\ref{eqn:YMFRD}).\footnote{This regularization seems similar to the 
one proposed in \cite{halp}. In a review article, they summarized 
extensive applications of their regularization scheme, based on the 
stochastic quantization, to a variety of quantum field theories.}




\section{Discussion}

As emphasized in the introduction, the cutoff scale $\Lambda$ of the 
gauge theory can be regarded as the radial scale $U$ in the AdS space, 
or more generally in the near horizon geometries of $Dp$-branes 
\cite{PP}.\footnote{Again we should be careful with the distinction 
between holographic and D-brane probes. See a footnote in the 
introduction.}
This simple but significant observation led us to contemplate 
the RG interpretation of the bulk/boundary duality, and pursue 
a gauge invariant formulation of the exact RG equation for the Yang-Mills 
theory. In particular to discuss the AdS/CFT duality, we apparently 
need to consider the supersymmetric extension of our formulation. 
For this purpose the analysis in \cite{WilAdS, OPE} and a supersymmetric 
Wilson loop in \cite{DGO} are certainly of importance. We would, 
however, like to discuss a possibility implied by our formulation 
in rather wider context of the holographic correspondence between the 
bulk gravity and the boundary gauge theory.

According to the wisdom of the AdS/CFT duality, a disturbance on the 
boundary will be responded by the bulk gravity as a gravitational fluctuation, 
in such a way that 
\begin{equation}
 \left\langle 
 \exp\left(\int_{\del M}d^D \vec{x} \phi_0(\vec{x} )
   {\cal O}(\vec{x} )\right) 
 \right\rangle
  = \exp\left(-S_{\mbox{grav}}[\phi (\vec{x} ,U)]\right),
 \label{eqn:CORRFNC}
\end{equation}
where $\del M$ is the boundary of the bulk space $M$, and 
${\cal O}(\vec{x} )$ is a local operator on the boundary field 
theory. Also $\phi(\vec{x} ,U)$ is a gravitational mode on the bulk 
space, and it becomes $\phi_0 (\vec{x} )$ at the boundary. Furthermore 
in the gravity action $S_{\mbox{grav}}$ the gravitational mode 
$\phi(\vec{x} ,U)$ is subject to the equation of motion, so that the 
r.h.s. depends only on the boundary value 
$\phi(\vec{x} ,\infty)=\phi_0 (\vec{x} )$ of a gravitational mode.

If we insist on a stronger conjecture \cite{Maldacena} on this duality 
that the string theory in the bulk is dual to the finite $N$ boundary 
gauge theory, we may consider the correlation functions of Wilson loops 
$W(C)$ instead of local operators ${\cal O}(\vec{x} )$. 
Then the gravity action might be replaced by the string field action. 
From this viewpoint we would like to recall our result (\ref{eqn:Hologram}) 
on the generating functional of Wilson loop correlators,
\begin{equation}
\left\langle 
 \exp\left(\sum_C J(C)W(C)\right) 
 \right\rangle =
Z[J]=e_D^{\widetilde{\calH} [{\delta \over \delta J},J;\Lambda]
     }Z_b[J].
  \label{eqn:SFVSFT}
\end{equation}
Now let us put the RG interpretation of the bulk/boundary duality 
into this argument. The bulk physics at some scale $U$ may be given 
by the boundary gauge theory in which degrees of freedom at higher 
momentum modes than the cutoff scale $\Lambda$ are integrated out. 
In this regard $Z_b[J]$ denotes the unintegrated form of the 
generating functional, and so it corresponds to the bulk physics at 
the IR limit or the boundary. Then the operator 
$\widetilde{\calH} [{\delta \over \delta J},J;\Lambda]$ gives the RG 
flow from UV to IR of the boundary gauge theory, that corresponds 
to a mapping from the boundary to the interior of the bulk gravity. 
In this sense we would like to regard the RG flow operator 
$\widetilde{\calH} [{\delta \over \delta J},J;\Lambda]$ as the holographic 
mapping of the boundary data. Moreover as we discussed in the last section, 
the RG equation implies ${d \over d\Lambda}Z[J]=0$, which is tantamount 
to a regularized version of the loop equation. Thus we might interpret 
the regularized loop equation as the string field equation of motion in 
the bulk space. 
Remember also that the RG flow operator 
$\widetilde{\calH} [{\delta \over \delta J},J;\Lambda]$ has a form 
of the string field Hamiltonian which consists of the terms that describe 
the joining and the splitting of strings and the string propagation.
These facts fit at least literally in a stronger conjecture on the 
bulk/boundary duality.

Apart from the RG interpretation of the bulk/boundary duality, we 
would like to mention a similarity of our formulation with the 
stochastic quantization of the Yang-Mills theory. In fact the authors in 
\cite{Periwal} proposed an interpretation of the Fokker-Planck 
Hamiltonian of the Yang-Mills theory as the string field Hamiltonian, 
initially as the one in the temporal gauge, just as discussed in the 
non-critical string theory, and later speculated an alternative 
interpretation that the fictitious time $\tau$ of the stochastic 
quantization can be thought of as the radial coordinate $U$ of the AdS 
space.\footnote{Actually the string field theory in 
the temporal gauge of non-critical strings was first proposed in 
\cite{IK}, and subsequently it was reconstructed in \cite{JR} as a 
collective field theory of stochastic quantization of matrix models 
in the double scaling limit. 
Also the intrinsic equivalence of the Fokker-Planck Hamiltonian and 
the loop operator was pointed out in \cite{Mar}.
The authors in \cite{Periwal} patched those ideas together, and 
added a new interpretation in the context of the Polyakov's non-critical 
strings \cite{Polyakov2} and also of the AdS/CFT duality.} Our 
formulation is close to their latter speculation. This similarity 
originates from the fact that the exact RG and the Fokker-Planck equations 
are quite similar diffusion equations with the cutoff $\Lambda$ and the 
fictitious time $\tau$ respectively as the time. It, however, seems 
hard to give a physical meaning to the finite value of the fictitious 
time $\tau$ in the stochastic quantization . 


\section*{Acknowledgment}


We would like to thank Hajime Aoki, Aki Hashimoto, 
Marty Halpern, Kentaro Hori, Mitsuhiro Kato, Joe Polchinski, 
Soo-Jong Rey and Rikard von Unge for useful discussions and comments. 
This work was supported in part by the Japan Society for the Promotion 
of Science and by the National Science Foundation under Grant No. PHY94-07194.




\section*{Appendix}
\setcounter{equation}{0}
\renewcommand{\theequation}{A.\arabic{equation}}

The computation of $\widetilde{\calH} [W,{\delta \over \delta W};\Lambda]$ can be done by mixture of the technique of the 
collective field method in \cite{JS} and that 
of \cite{Polyakov} presented in a derivation of the loop equation.
In terms of the gauge fields $A^{\mu}(\vec{x} )$, it is expressed as 
\begin{eqnarray}
\widetilde{\calH} [W,{\delta \over \delta W};\Lambda]\!\!\!
 &=& 
 \!\!\!{1 \over 2N}
 \int d^D \vec{x} \int d^D \!\!\vec{y} 
 \int \!\!d^D \vec{p} {1 \over p^2}e^{i\vec{p} \cdot(\vec{x-y} )}
 \Biggl[ \nn\\
 &&\left. \!\!\!\!\! - K^{'}(p^2/\Lambda^2)
 \Tr\left(T^b \Phi[\Gamma_{\vec{x} \vec{y} }]T^a
 \Phi^{-1}[\Gamma_{\vec{x} \vec{y} }]\right)
 \frac{\delta^2}{\delta A^{\mu}_a(\vec{y} )\delta A^{\mu}_b(\vec{x} )}
 \right. \label{eqn:tildeH}\\
 && \!\!\!\!\! -\left. K^{'}(p^2/\Lambda^2)K^{-1}(p^2/\Lambda^2)\Tr\left(T^b
 \Phi[\Gamma_{\vec{x} \vec{y} }]{1 \over g_b^2}D^{\nu}F_{\nu\mu}(\vec{y} )
 \Phi^{-1}[\Gamma_{\vec{x} \vec{y} }]\right)
 \frac{\delta}{\delta A^{\mu}_b (\vec{x} )}\right].\nn
\end{eqnarray}
The second derivative term consists of two pieces, 
the joining and the splitting of strings, due to the chain rule, 
as in (the first paper of) \cite{JS}. The one that describes the 
joining of strings is 
\begin{equation}
\sum_{C,C^{\prime}}
\Tr\left(T^b \Phi[\Gamma_{\vec{x} \vec{y} }]T^a
 \Phi^{-1}[\Gamma_{\vec{x} \vec{y} }]\right)
 \frac{\delta W(C)}{\delta A^{\mu}_a (\vec{y} )}
 \frac{\delta W(C^{\prime})}{\delta A^{\mu}_b (\vec{x} )}
 \frac{\delta}{\delta W(C^{\prime})}\frac{\delta}{\delta W(C)}.
 \label{eqn:joining}
\end{equation}
Actually the joining property can be easily understood from an 
explicit calculation. 
\begin{eqnarray}
N^2\Tr\left(T^b \Phi[\Gamma_{\vec{x} \vec{y} }]T^a
 \Phi^{-1}[\Gamma_{\vec{x} \vec{y} }]\right)
 \frac{\delta W(C)}{\delta A^{\mu}_a (\vec{y} )}
 \frac{\delta W(C^{\prime})}{\delta A^{\mu}_b (\vec{x} )} 
  \nspace\nspace\nspace\nspace\nspace && \nn\\
 &=&-\int_0^{2\pi} ds \int_0^{2\pi} d\esprime 
 \left(\vec{\dot{x}} (s)\cdot\vec{\dot{x}} (\esprime)\right)
 \delta^D \left(\vec{x} (s)-\vec{y} \right)
 \delta^D \left(\vec{x} (\esprime)-\vec{x} \right) \nn\\
 &\times&
 \Tr\left[\left( Pe^{i\int_0^{s}d\esdbprime\vec{\dot{x}} 
 (\esdbprime) \cdot\vec{A} (\vec{x} (\esdbprime)) }\right)
 \Phi^{-1}[\Gamma_{\vec{x} \vec{y} }]
 \left( Pe^{i\int_{\esprime}^{2\pi}d\esdbprime\vec{\dot{x}} 
 (\esdbprime) \cdot\vec{A} (\vec{x} (\esdbprime)) }\right) 
 \right. \nn\\
 &\times&\left. 
 \left( Pe^{i\int_0^{\esprime}d\esdbprime\vec{\dot{x}} (\esdbprime) 
 \cdot\vec{A} (\vec{x} (\esdbprime)) }\right)\Phi [\Gamma_{\vec{x} \vec{y} }]
 \left( Pe^{i\int_{s}^{2\pi}d\esdbprime\vec{\dot{x}} (\esdbprime) 
 \cdot\vec{A} (\vec{x} (\esdbprime)) }\right)
 \right] 
 \label{eqn:compjoin}\\
 &=&
 -N\int_0^{2\pi} ds \int_0^{2\pi} d\esprime 
 \left(\vec{\dot{x}} (s)\cdot\vec{\dot{x}} (\esprime)\right)
 \delta^D \left(\vec{x} (s)-\vec{y} \right)
 \delta^D \left(\vec{x} (\esprime)-\vec{x} \right)
 W(C_M), \nn
\end{eqnarray}
where a loop $C_M$ denotes the merging of two loops $C$ and 
$C^{\prime}$, and it is composed of four lines, i.e., 
$C_M=C_{\vec{x} (s)\vec{x} (s)}\Gamma_{\vec{y} \vec{x} }
C_{\vec{x} (\esprime)\vec{x} (\esprime)}\Gamma_{\vec{x} \vec{y} }$.
Here we define the product of line contours as an oriented contour 
in which each line is connected at a common point. Also we used different 
symbols $C_{\vec{x} \vec{y} }$ and $\Gamma_{\vec{x} \vec{y} }$ for line 
contours, in order to distinguish a pair of lines 
$\Gamma_{\vec{x} \vec{y} }$ introduced in the field redefinition 
(\ref{eqn:YMFRD}) from the other lines $C_{\vec{x} \vec{y} }$ 
which are segments of loops $C$ and $C^{\prime}$.

The part corresponding to the splitting of strings is given by 
\begin{equation}
\sum_{C}
\Tr\left(T^b \Phi[\Gamma_{\vec{x} \vec{y} }]T^a
 \Phi^{-1}[\Gamma_{\vec{x} \vec{y} }]\right)
 \frac{\delta^2 W(C)}{\delta A^{\mu}_b (\vec{x} )
 \delta A^{\mu}_a (\vec{y} )}
 \frac{\delta}{\delta W(C)}.
 \label{eqn:split}
\end{equation}
Similarly it can be rewritten in terms of Wilson loops as
\begin{eqnarray}
N\Tr\left(T^b \Phi[\Gamma_{\vec{x} \vec{y} }]T^a
 \Phi^{-1}[\Gamma_{\vec{x} \vec{y} }]\right)
 \frac{\delta^2 W(C)}{\delta A^{\mu}_b (\vec{x} )
 \delta A^{\mu}_a (\vec{y} )}
 \nspace\nspace\nspace\nspace\nspace && \nn\\
 &=& -\int_0^{2\pi} ds\left[
 \int_0^s d\esprime\Tr\left\{ 
 \left( Pe^{i\int_0^{\esprime} d\esdbprime\vec{\dot{x}} (\esdbprime) 
 \cdot\vec{A} (\vec{x} (\esdbprime)) }\right)\Phi [\Gamma_{\vec{x} \vec{y} }]
 \left( Pe^{i\int_{s}^{2\pi}d\esdbprime\vec{\dot{x}} (\esdbprime) 
 \cdot\vec{A} (\vec{x} (\esdbprime)) }\right)\right\}\right. \nn\\
 &&\bigquad\qquad\quad\times
 \Tr\left\{ 
 \left( Pe^{i\int_{\esprime}^s d\esdbprime\vec{\dot{x}} (\esdbprime) 
 \cdot\vec{A} (\vec{x} (\esdbprime)) }\right)
 \Phi^{-1} [\Gamma_{\vec{x} \vec{y} }]\right\}\nn\\
 &+& 
 \int_s^{2\pi} d\esprime\Tr\left\{ 
 \left( Pe^{i\int_0^s d\esdbprime\vec{\dot{x}} (\esdbprime) 
 \cdot\vec{A} (\vec{x} (\esdbprime)) }\right)\Phi^{-1} [\Gamma_{\vec{x} \vec{y} }]
 \left( Pe^{i\int_{\esprime}^{2\pi}d\esdbprime\vec{\dot{x}} (\esdbprime) 
 \cdot\vec{A} (\vec{x} (\esdbprime)) }\right)\right\} \nn\\
 &&\left. \bigquad\qquad\quad\times
 \Tr\left\{ 
 \left( Pe^{i\int_s^{\esprime} d\esdbprime\vec{\dot{x}} (\esdbprime) 
 \cdot\vec{A} (\vec{x} (\esdbprime)) }\right)
 \Phi [\Gamma_{\vec{x} \vec{y} }]\right\} \right]\nn\\
 &&\times
 \left(\vec{\dot{x}} (s)\cdot\vec{\dot{x}} (\esprime)\right)
 \delta^D \left(\vec{x} (s)-\vec{y} \right)
 \delta^D \left(\vec{x} (\esprime)-\vec{x} \right)
  \label{eqn:cmpsplit}\\
 &=&
 -N^2\int_0^{2\pi} ds \int_0^{2\pi} d\esprime 
 \left(\vec{\dot{x}} (s)\cdot\vec{\dot{x}} (\esprime)\right)
 \delta^D \left(\vec{x} (s)-\vec{y} \right)
 \delta^D \left(\vec{x} (\esprime)-\vec{x} \right)
 W(C_B)W(C_{{\bar B}}),\nn
\end{eqnarray}
where a loop $C$ is broken into two loops $C_B$ and $C_{{\bar B}}$, 
and they are respectively given by $C_B=\Gamma_{\vec{x} \vec{y} }
C_{\vec{x} (s)\vec{x} (\esprime)}$ and $C_{{\bar B}}=C_{\vec{x} 
(\esprime)\vec{x} (s)}\Gamma_{\vec{y} \vec{x} }$.

Finally the second term in eq.(\ref{eqn:tildeH}) corresponds to the 
kinetic term of the string field, 
\begin{eqnarray}
\sum_C \Tr\left(T^b
 \Phi[\Gamma_{\vec{x} \vec{y} }]{1 \over g_b^2}D^{\nu}F_{\nu\mu}(\vec{y} )
 \Phi^{-1}[\Gamma_{\vec{x} \vec{y} }]\right)
 \frac{\delta W(C)}{\delta A^{\mu}_b (\vec{x} )}
 \frac{\delta}{\delta W(C)}
 \nspace\nspace\nspace\nspace\nspace\nspace && \nn\\
 &=&
 \sum_C {i \over g_b^2 N}
 \int_0^{2\pi} ds \dot{x}^{\mu}(s)\delta^D (\vec{x} (s)-\vec{x} )
 \nn\\
  &\!\!\!\!\!\!\!\times&\!\!\!\!\!\!\!\!
 \Tr\left\{ \left( Pe^{i\int_0^s
 d\esprime\vec{\dot{x}} (\esprime) 
 \cdot\vec{A} (\vec{x} (\esprime)) }\right)\Phi [\Gamma_{\vec{x} \vec{y} }]
 D^{\nu}F_{\nu\mu}(\vec{y} )\Phi^{-1} [\Gamma_{\vec{x} \vec{y} }]
 \left( Pe^{i\int_s^{2\pi}d\esprime\vec{\dot{x}} (\esprime) 
 \cdot\vec{A} (\vec{x} (\esprime)) }\right)
 \right\} \frac{\delta}{\delta W(C)}\nn\\
 &=&
 {1 \over g_b^2}\int_0^{2\pi} ds \delta^D (\vec{x} (s)-\vec{x} )
 \left. \frac{\delta^2 W(C)}{\delta x(s_0)^2}
 \right|_{\vec{x} (s_0)=\vec{y} }\frac{\delta}{\delta W(C)},
 \label{eqn:DFMLP}
\end{eqnarray}
where we introduced a local derivative of the loop space \cite{Polyakov}, 
\begin{equation}
\frac{\delta^2}{\delta x(s)^2}=\lim_{\epsilon\to 0}
\int_{-\epsilon}^{\epsilon} dt
\frac{\delta^2}{\delta x^{\mu}(s+t/2) \delta x_{\mu}(s-t/2)}.
\label{eqn:localdv}
\end{equation}
%





\begin{thebibliography}
%
\bibitem{SS} N.~Seiberg, 
         \lq\lq Why Is The Matrix Model Correct?,"
         \prl {\bf 79} (1997) 3577, hep-th/9710009; \\
         A.~Sen,
         \lq\lq D0-Branes On $T^{n}$ And Matrix Theory,"
         Adv. Theor. Math. Phys. {\bf 2} (1998) 51, 
         hep-th/9709220.
\bibitem{BFSS} T.~Banks, W.~Fishler, S.~Shenker and L.~Susskind,
         \lq\lq M-Theory As A Matrix Model: A Conjecture," 
         \pr {\bf D55} (1997) 5112, hep-th/9610043; \\ 
         L.~Susskind,
         \lq\lq Another Conjecture About M(atrix) Theory,"
         hep-th/9704080.
\bibitem{Maldacena} J.~Maldacena,
         \lq\lq The Large $N$ Limit Of Superconformal Field Theories 
                And Supergravity," 
          Adv. Theor. Math. Phys. {\bf 2} (1998) 231, 
          hep-th/9711200; \\
         See also for a review, O. Aharony, S. S. Gubser, J. Maldacena, 
           H. Ooguri and Y. Oz, 
          \lq\lq Large $N$ Field Theories, String Theory And 
                 Gravity,"
          hep-th/9905111.
\bibitem{IMSY} N.~Itzhaki, J.~Maldacena, J.~Sonnenschein and S.~Yankielowicz,
         \lq\lq Supergravity And The Large $N$ Limit Of Theories 
                With Sixteen Supercharges,"
         \pr {\bf D58} (1998) 46004, hep-th/9802042.
\bibitem{TASI} J.~Polchinski, 
         \lq\lq Dirichlet-Branes And Ramond-Ramond Charges,"
         \prl {\bf 75} (1995) 4724, hep-th/9510017; 
         See also for a review, 
         \lq\lq TASI Lectures On D-branes,"
         hep-th/9611050.
\bibitem{GKP} S.~S.~Gubser, I.~R.~Klebanov and A.~M.~Polyakov,
         \lq\lq Gauge Theory Correlators From Non-critical String Theory,"
         \pl {\bf B428} (1998) 105, hep-th/9802109.
\bibitem{Witten} E.~Witten,
         \lq\lq Anti-de Sitter Space And Holography,"
         Adv. Theor. Math. Phys. {\bf 2} (1998) 253, 
         hep-th/9802150.
\bibitem{SW} L.~Susskind and E.~Witten,
	\lq\lq The Holographic Bound in Anti-de Sitter Space,''
	     hep-th/9805114.
\bibitem{PP} A.~W.~Peet and J.~Polchinski,
        \lq\lq UV/IR Relations In AdS Dynamics,"
        \pr {\bf D59} (1999) 065011, hep-th/9809022.
\bibitem{WK} K. G. Wilson and J. Kogut,
        \lq\lq The Renormalization Group And The $\epsilon$
               Expansion,"
        Phys. Rep. {\bf 12C} (1974) 75.
\bibitem{JS} B. Sakita,
        \lq\lq Field Theory Of Strings As A Collective Field 
               Theory Of U($N$) Gauge Fields,"
        \pr {\bf D21} (1980) 1067; \\
             A. Jevicki and B. Sakita,
        \lq\lq The Quantum Collective Field Method And Its 
               Application To The Planar Limit,"
        \np {\bf B165} (1980) 511.
\bibitem{AdSRG} E. T. Akhmedov,
        \lq\lq A Remark On The AdS/CFT Correspondence And
               The Renormalization Group Flow,"
        \pl {\bf B442} (1998) 152, hep-th/9806217; \\
                E. Alvarez and C. Gomez, 
        \lq\lq Geometric Holography, The Renormalization Group
               And The c-Theorem," 
        \np {\bf B541} (1999) 441, hep-th/9807226; \\
                M. Porrati and A. Starinets, 
        \lq\lq RG Fixed Points In Supergravity Duals of 4-d 
               Field Theory And Asymptotically AdS Spaces," 
        \pl {\bf B454} (1999) 77, hep-th/9903085; \\
                V. Balasubramanian and P. Kraus,
        \lq\lq Spacetime And The Holographic Renormalization 
               Group," 
        hep-th/9903190; \\
                H. Verlinde,
        \lq\lq Holography and Compactification,"
        hep-th/990618; \\
                O. Andreev,
        \lq\lq Probing $\mbox{AdS}_3$/CFT Correspondence 
               Via World-Sheet Methods And 2d Gravity 
               Like Scaling Arguments,"
        hep-th/9909222.
\bibitem{SJR} S.-J. Rey, private communication.
\bibitem{Morris} T. R. Morris,
        \lq\lq A Manifestly Gauge Invariant Exact 
               Renormalization Group,"
        hep-th/9810104; 
        \lq\lq A Gauge Invariant Exact Renormalization
               Group I,"
        hep-th/9910058.
\bibitem{Polchinski} J. Polchinski,
        \lq\lq Renormalization And Effective Lagrangian,"
        \np {\bf B231} (1984) 269.
\bibitem{MM} Y. M. Makeenko and A. A. Migdal,
        \lq\lq Exact Equation For The Loop Average In 
               Multicolor QCD,"
        \pl {\bf 88B} (1979) 135;
        \lq\lq Quantum Chromodynamics As Dynamics Of Loops,"
        \np {\bf B188} (1981) 269.
\bibitem{Polyakov} A. M. Polyakov,
        \lq\lq Gauge Fields and Strings,"
        (1987), Harwood Academic Publishers.
\bibitem{halp} M. B. Halpern and Y. Makeenko, 
        \lq\lq Continuum Regularized Loop Space Equation,"
        \pl {\bf B218} (1989) 230; 
        See also for a review, 
        M. B. Halpern, 
        \lq\lq Universal Continuum Regularization Of 
               Quantum Field Theory,"
        \ptp Suppl. {\bf 111} (1993) 163.
\bibitem{WilAdS} J. M. Maldacena,
        \lq\lq Wilson Loops In Large $N$ Field Theories,"
        \prl {\bf 80} (1998) 4859, hep-th/9803002; \\
              S.-J. Rey and J. Yee,
        \lq\lq Macroscopic Strings As Heavy Quarks Of 
               Large $N$ Gauge Theory And Anti-de Sitter 
               Supergravity," 
        hep-th/9803001.
\bibitem{OPE} D. Berenstein, R. Corrado, W. Fischler and 
              J. Maldacena,
        \lq\lq The Operator Product Expansion For Wilson 
               Loops And Surfaces In The Large $N$ Limit,"
        \pr {\bf D59} (1999) 105023, hep-th/9809188.
\bibitem{DGO} N. Drukker, D. J. Gross and H. Ooguri,
        \lq\lq Wilson Loops And Minimal Surfaces,"
        hep-th/9904191; \\
              N. Drukker,
        \lq\lq A New Type Of Loop Equations,"
        hep-th/9908113.
\bibitem{Periwal} V. Periwal, 
        \lq\lq String Field Theory Hamiltonians From Yang-Mills 
               Theories,"
        hep-th/9906052; 
        \lq\lq A Toy Model Of Polyakov Duality,"
        hep-th/9908203; \\
             G. Lifschytz and V. Periwal,
        \lq\lq Dynamical Truncation Of The String Spectrum At 
               Finite $N$,"
        hep-th/9909152.
\bibitem{IK} N. Ishibashi and H. Kawai,
        \lq\lq String Field Theory Of Noncritical Strings,"
        \pl {\bf B314} (1993) 190, hep-th/9307045; \\
        See also for a review, 
        M. Ikehara, N. Ishibashi, H. Kawai, T. Mogami, 
        R. Nakayama and N. Sasakura, 
        \lq\lq A Note On String Field Theory In The 
               Temporal Gauge,"
        \ptp Suppl. {\bf 118} (1995) 241, hep-th/9409101.
\bibitem{JR} A. Jevicki and J. Rodrigues,
        \lq\lq Loop Space Hamiltonians And Field Theory 
               Of Noncritical Strings,"
        \np {\bf B421} (1994) 278, hep-th/9312118.
\bibitem{Mar} G. Marchesini, 
        \lq\lq A Comment On The Stochastic Quantization: 
               The Loop Equation Of The Gauge Theory As 
               The Equilibrium Condition,"
        \np {\bf B191} (1981) 214.
\bibitem{Polyakov2} A. M. Polyakov,
        \lq\lq String Theory And Quark Confinement,"
        \np Proc. Suppl. {\bf 68} (1998) 1, hep-th/9711002.
%
\end{thebibliography}
\end{document}